\documentclass[a4paper]{jpconf}
\usepackage{graphicx}
\usepackage{amsmath,amssymb,bm}

\def \beq {\begin{equation}}
\def \edq {\end{equation}}
\def \bes {\begin{subequations}}
\def \eds {\end{subequations}}
\def \beqn {\begin{equation*}}
\def \edqn {\end{equation*}}

\def \dag {\dagger}

\def \up {\uparrow}
\def \down {\downarrow}

\def \calj {{\cal{J}}}

\begin{document}
\title{Thermoelectric effects in quantum Hall systems beyond linear response}

\author{Rosa L\'{o}pez$^1$, Sun-Yong Hwang$^1$, David S\'{a}nchez$^1$}

\address{$^1$ Institut de F\'{\i}sica Interdisciplin\`aria i de Sistemes Complexos
IFISC (CSIC-UIB), E-07122 Palma de Mallorca, Spain}

\ead{rosa.lopez-gonzalo@uib.es}

\begin{abstract} We consider a quantum Hall system with an antidot acting as an energy dependent
scatterer. In the purely charge case, we find deviations from the Wiedemann-Franz law that take
place in the nonlinear regime of transport. We also discuss Peltier effects beyond linear response
and describe both effects using magnetic-field asymmetric transport coefficients.
For the spin case such as that arising along the helical edge states of a two-dimensional
topological insulator, we investigate the generation of spin currents as a result of applied
voltage and temperature differences in samples attached to ferromagnetic leads.
We find that in the parallel configuration the spin current can be tuned with the leads'
polarization even in the linear regime of transport. In contrast, for antiparallel magnetizations the
spin currents has a strict nonlinear dependence on the applied fields.
\end{abstract}

\section{Introduction}
Under the application of strong fields, electron transport becomes nonlinear and new effects arise: rectification \cite{son98,lin00,sho01,fle02,but03,gon04,hac04}, magnetic-field asymmetries\cite{san04,spi04,mart06,and06,kal09,her09,lim10,kub11,lud13,wei06,mar06,let06,zum06,ang07,har08,che09,bra09}, and the generation of higher harmonics \cite{ver93,Guo00,Safi11}.  Quantum coherent conductors are excellent platforms  to observe and study nonlinearities since small voltage biases cause  a  sizeable effect over rather short lengths. In the ballistic regime of transport, current response is determined by the transmission probability $t(E)$. When $t$ depends weakly on the carrier's energy $E$, current is always a linear function of $V$ independently of the background temperature $T$. Quite generally, however, transmission across nanostructures shows sharp features arising from strongly energy dependent $t(E)$. Furthermore, in the nonlinear regime of transport, the transmission becomes a function of the applied voltage $V$ because the applied field modifies the potential landscape which in turn affects the scattering properties of the sample \cite{but93,chr96}. The self-consistent procedure must then include electron-electron interactions that restore current conservation and gauge invariance beyond linear response. In fact, it is the nontrivial dependence of the screening potential on voltage and magnetic fields that explains both rectification effects and magnetic-asymmetries, respectively.

Electron motion can also be induced with the application of external thermal gradients. The two main thermoelectric effects are the Seebeck and Peltier effects. The former leads to the creation of a voltage drop in response to a temperature difference $\theta$ in the open circuit case. The latter is based on the fact that electrons carry energy in addition to charge, and a heat current then flows in the presence of an electric current. Both phenomena have been observed in mesoscopic systems \cite{mol92,dzu97,god99}, and the agreement with the scattering approach \cite{but90} is remarkable. In these nanodevices, the thermodynamic efficiency can be tuned with an external magnetic field \cite{sai11,ent12,ben11,bra13,bal13,ape13}.

Beyond linear response, one must also take into account that the transmission is not only a function of $E$ and $V$ but also depends on $\theta$: $t(E,V,\theta)$ \cite{lop13a}. As a consequence, the injected charge that builds up in the vicinity of the sample is determined from both particle and entropic injectivities \cite{lop13a,mea13,lop13b}. Upper bounds to the performance of heat engines and coolers are thus to be carefully evaluated \cite{whi13a,whi13b}. Surprisingly, the thermovoltage can vanish for a finite value of $\theta$, as recently demonstrated in experiments with nanowire quantum dots \cite{fah13}. Furthermore, reciprocal relations are shown to break down for sufficiently high heating currents applied to multiterminal setups \cite{mat12}.  Interestingly, departures from the Onsager-Kelvin relations differ from auto- or cross-terminal measurements \cite{lop13a,hwa13}.

In this work, we discuss thermoelectric effects beyond linear response for quantum Hall bars with an inserted antidot. This system allows for an investigation of the underlying symmetries while providing, at the same time, a strong interaction and energy dependent scattering \cite{for94,kir94}. We find large deviations of the Wiedemann-Franz law and interesting nonlinear behavior of the Peltier effect. Additionally, we discuss the spin transport in quantum spin Hall systems \cite{QiZhang} coupled to ferromagnetic leads. Whereas the normal case already leads to the generation of pure spin currents \cite{hwa14}, we here find a competition with the polarized currents injected from the magnetic contacts.

\section{Theoretical formalism}
	When a mesoscopic conductor is coupled to multiple terminals $\alpha,\beta,\dots$, each terminal can be characterized by voltage bias $eV_{\alpha}=\mu_{\alpha}-E_{F}$ ($\mu_{\alpha}$ is the electrochemical potential and $E_{F}$ is the Fermi energy) and also by the temperature gradient $\theta_{\alpha}=T_{\alpha}-T$ ($T_{\alpha}$ and $T$ are the reservoir and the background temperature, respectively).
	In the presence of these two driving fields, the electronic and heat transport is entirely described by the scattering matrix $s_{\alpha\beta}=s_{\alpha\beta}(E, eU)$, which is a function of the carrier energy $E$ and the potential landscape $U$ inside the conductor.
	The potential $U=U(\vec{r},\{V_{\gamma}\},\{\theta_{\gamma}\})$ is, in general, a function of the position $\vec{r}$ and the set of applied voltages $\{V_{\gamma}\}$ and thermal gradients $\{\theta_{\gamma}\}$.
	The charge and heat currents at lead $\alpha$ are respectively given by~\cite{lop13a,lop13b}
\begin{align}
I_{\alpha}&=\frac{2e}{h}\sum_{\beta}\int dE A_{\alpha\beta}(E,eU)f_{\beta}(E),\label{I}\\
{\cal J}_{\alpha}&=\frac{2}{h}\sum_{\beta}\int dE(E-\mu_{\alpha}) A_{\alpha\beta}(E,eU)f_{\beta}(E),\label{J}
\end{align}
where $A_{\alpha\beta}=\text{Tr}[\delta_{\alpha\beta}-s_{\alpha\beta}^{\dag}s_{\alpha\beta}]$ and $f_{\beta}(E)=(1+\exp[(E-\mu_{\beta})/k_{B}T_{\beta}])^{-1}$ is the Fermi distribution function in the reservoir $\beta$.
	In the weakly nonlinear regime of transport, one can expand these currents around the equilibrium state, i.e., $\mu_{\alpha}=E_{F}$ and $T_{\alpha}=T$, up to second order of $V_{\alpha}$ and $\theta_{\alpha}$~\cite{lop13a,mea13,lop13b}:
\begin{align}
I_{\alpha}&=\sum_{\beta}\Big(G_{\alpha\beta}V_{\beta}+L_{\alpha\beta}\theta_{\beta}\Big)
	+\sum_{\beta\gamma}\Big(G_{\alpha\beta\gamma}V_{\beta}V_{\gamma}
	+L_{\alpha\beta\gamma}\theta_{\beta}\theta_{\gamma}
	+2M_{\alpha\beta\gamma}V_{\beta}\theta_{\gamma}\Big),\label{elec}\\
{\cal J}_{\alpha}&=\sum_{\beta}\Big(R_{\alpha\beta}V_{\beta}+K_{\alpha\beta}\theta_{\beta}\Big)
	+\sum_{\beta\gamma}\Big(R_{\alpha\beta\gamma}V_{\beta}V_{\gamma}
	+K_{\alpha\beta\gamma}\theta_{\beta}\theta_{\gamma}+2H_{\alpha\beta\gamma}V_{\beta}\theta_{\gamma}\Big).\label{heat}
\end{align}
	The linear conductance coefficients are given by~\cite{but90}
$G_{\alpha\beta}=(2e^{2}/h)\int dE A_{\alpha\beta}(E)(-\partial_{E}f)\approx(2e^{2}/h)A_{\alpha\beta}(E_{F})$,
$L_{\alpha\beta}=(2e/hT)\int dE(E-E_{F})A_{\alpha\beta}(E)(-\partial_{E}f)\approx(2e\pi^{2}k_{B}^{2}T/3h)\partial_{E}A_{\alpha\beta}(E)|_{E=E_{F}}$,
$R_{\alpha\beta}=(2e/h)\int dE(E-E_{F}) A_{\alpha\beta}(E)(-\partial_{E}f)\approx(2e\pi^{2}k_{B}^{2}T^{2}/3h)\partial_{E}A_{\alpha\beta}(E)|_{E=E_{F}}$, and
$K_{\alpha\beta}=(2/h)\int dE\frac{(E-E_{F})^{2}}{T} A_{\alpha\beta}(E)(-\partial_{E}f)\approx(2\pi^{2}k_{B}^{2}T/3h)A_{\alpha\beta}(E_{F})$, where we have resorted to the Sommerfeld expansion for the approximate expressions.
	We emphasize here that the linear coefficients $G_{\alpha\beta}$, $L_{\alpha\beta}$, $R_{\alpha\beta}$, and $K_{\alpha\beta}$ are equilibrium quantities and hence are independent of the screening potential $U$.
	In contrast, the nonlinear coefficients $G_{\alpha\beta\gamma}$, $L_{\alpha\beta\gamma}$, $M_{\alpha\beta\gamma}$, $R_{\alpha\beta\gamma}$, $K_{\alpha\beta\gamma}$, and $H_{\alpha\beta\gamma}$ manifestly depend on $U$ in response to the applied voltage and temperature biases. Explicitly, they are written by~\cite{lop13a,mea13,lop13b}
$G_{\alpha\beta\gamma}=(-e^{2}/h)\int dE\Big(\frac{\partial A_{\alpha\beta}}{\partial V_{\gamma}}+\frac{\partial A_{\alpha\gamma}}{\partial V_{\beta}}+e\delta_{\beta\gamma}\frac{\partial A_{\alpha\beta}}{\partial E}\Big)\partial_{E}f$,
$L_{\alpha\beta\gamma}=(e/h)\int dE\frac{E_{F}-E}{T}\Big(\frac{\partial A_{\alpha\beta}}{\partial \theta_{\gamma}}+\frac{\partial A_{\alpha\gamma}}{\partial \theta_{\beta}}+\delta_{\beta\gamma}\frac{E-E_{F}}{T}\frac{\partial A_{\alpha\beta}}{\partial E}\Big)\partial_{E}f$,
$M_{\alpha\beta\gamma}=(e^{2}/h)\int dE\Big(\frac{E_{F}-E}{eT}\frac{\partial A_{\alpha\gamma}}{\partial V_{\beta}}-\frac{\partial A_{\alpha\beta}}{\partial \theta_{\gamma}}-\delta_{\beta\gamma}\frac{E-E_{F}}{T}\frac{\partial A_{\alpha\beta}}{\partial E}\Big)\partial_{E}f$,
$R_{\alpha\beta\gamma}=(e^{2}/h)\int dE\Big\{\delta_{\alpha\gamma}A_{\alpha\beta}+\delta_{\alpha\beta}A_{\alpha\beta}-(E-E_{F})\Big(\frac{\partial A_{\alpha\beta}}{\partial eV_{\gamma}}+\frac{\partial A_{\alpha\gamma}}{\partial eV_{\beta}}\Big)-\delta_{\beta\gamma}\Big[(E-E_{F})\frac{\partial A_{\alpha\beta}}{\partial E}+A_{\alpha\beta}\Big]\Big\}\partial_{E}f$,
$K_{\alpha\beta\gamma}=(-1/h)\int dE\frac{(E-E_{F})^{2}}{T}\Big\{\Big(\frac{\partial A_{\alpha\beta}}{\partial \theta_{\gamma}}
			+\frac{\partial A_{\alpha\gamma}}{\partial \theta_{\beta}}\Big)
			+\delta_{\beta\gamma}\Big[\frac{(E-E_{F})}{T}\frac{\partial A_{\alpha\beta}}{\partial E}
			+\frac{A_{\alpha\beta}}{T}\Big]\Big\}\partial_{E}f$, and
$H_{\alpha\beta\gamma}=(-e/h)\int dE(E-E_{F})\Big\{
		\Big(\frac{\partial A_{\alpha\gamma}}{\partial \theta_{\beta}}
			+\frac{(E-E_{F})}{T}\frac{\partial A_{\alpha\beta}}{\partial eV_{\gamma}}
			-\delta_{\alpha\gamma}\frac{A_{\alpha\beta}}{T}\Big)
			+\delta_{\beta\gamma}\Big[\frac{(E-E_{F})}{T}\frac{\partial A_{\alpha\beta}}{\partial E}
			+\frac{A_{\alpha\beta}}{T}\Big]\Big\}\partial_{E}f$.

	Leading order interaction effects can be incorporated into the nonequilibrium potential $U$ via characteristic potentials (CPs) $u_{\alpha}=(\partial U/\partial V_{\alpha})_{\text{eq}}$ and $z_{\alpha}=(\partial U/\partial\theta_{\alpha})_{\text{eq}}$:
\begin{equation}\label{eq:U1}
U=U_{\text{eq}}+\sum_{\alpha}u_{\alpha}V_{\alpha}+\sum_{\alpha}z_{\alpha}\theta_{\alpha}.
\end{equation}
Note that CPs $u_{\alpha}$ and $z_{\alpha}$ relate the variation of $U$ to voltage and temperature shifts, respectively.
	We determine $U$ self-consistently by considering the net charge of the system $q=q_{\text{bare}}+q_{\text{scr}}$.
	The bare charge $q_{\text{bare}}=e\sum_{\alpha}(D_{\alpha}^{p}eV_{\alpha}+D_{\alpha}^{e}\theta_{\alpha})$ has two contributions due to voltage bias and temperature shift, described by the particle injectivity~\cite{but93,chr96} $\nu_{\alpha}^{p}(E)=(2\pi i)^{-1}\sum_{\beta}\text{Tr}\big[s_{\beta\alpha}^{\dag}\frac{ds_{\beta\alpha}}{dE}\big]$ and the entropic injectivity~\cite{lop13a} $\nu_{\alpha}^{e}(E)=(2\pi i)^{-1}\sum_{\beta}\text{Tr}\big[\frac{E-E_{F}}{T}s_{\beta\alpha}^{\dag}\frac{ds_{\beta\alpha}}{dE}\big]$ with $D_{\alpha}^{p,e}=-\int dE\nu_{\alpha}^{p,e}(E)\partial_{E}f$.
	The screening charge $q_{\text{scr}}$ building up inside the sample can be obtained from the response of the screening potential, $\Delta U=U-U_{\text{eq}}$, away from the equilibrium state $U_{\text{eq}}$.
	One can write $q_{\text{scr}}=e^{2}\Pi\Delta U$ where $\Pi$ is the Lindhard function which becomes $\Pi=\int dED(E)\partial_{E}f$ in the long wavelength approximation, with $D(E)$ being the density of states.
	Then, the set of equations for the CPs is closed via the Poisson's equation $\nabla^{2}\Delta U=-4\pi q$.

	Finally, for a practical calculation, one can resort to the WKB approximation valid in the long wavelength limit and make the replacement $\delta/\delta U\to-e\partial/\partial E$, i.e., $\frac{\partial A_{\alpha\beta}}{\partial V_{\gamma}}=\frac{\partial U}{\partial V_{\gamma}}\frac{\delta A_{\alpha\beta}}{\delta U}\approx-eu_{\gamma}\frac{\partial A_{\alpha\beta}}{\partial E}$ and $\frac{\partial A_{\alpha\beta}}{\partial\theta_{\gamma}}=\frac{\partial U}{\partial\theta_{\gamma}}\frac{\delta A_{\alpha\beta}}{\delta U}\approx-ez_{\gamma}\frac{\partial A_{\alpha\beta}}{\partial E}$.
	For a two-terminal setup which we consider below, the matrix elements of $A_{\alpha\beta}$ are given by $A_{11}=A_{22}=-A_{12}=-A_{21}=t(E)$, where $t(E)$ is the transmission probability through the system.

\subsection{Magnetic-field asymmetry}
	At equilibrium, the screening potential $U=U_{\text{eq}}$ is symmetric with respect to the reversal of an applied magnetic field, i.e., $U_{\text{eq}}(B)=U_{\text{eq}}(-B)$, due to the fundamental microscopic reversibility principle.
	In Ref.~\cite{but90}, magnetic-field \emph{symmetry} of \emph{linear} thermoelectric and heat transport has been shown in the framework of scattering theory.
	When the system is driven into the nonequilibrium regime, however, this magnetic-field symmetry can be broken since the CPs in Eq.~\eqref{eq:U1} are in general magnetic-field asymmetric, i.e., $u_{\alpha}(B)\ne u_{\alpha}(-B)$ and $z_{\alpha}(B)\ne z_{\alpha}(-B)$.
	In Ref.~\cite{san04}, the nonlinear conductance $G_{\alpha\beta\gamma}$ in the isothermal case has been shown to be magnetic-field asymmetric since $u_{\alpha}$, the voltage response of $U$, is not an even function of $B$.
	Here, we generalize this result to all thermoelectric and heat transport coefficients.

	We quantify the magnetic-field asymmetry in the nonlinear transport regime by defining the symmetry($\Sigma$) and the asymmetry($A$) parameters~\cite{hwa13} for $G$, $L$, $R$, and $K$ coefficients in Eqs.~\eqref{elec} and~\eqref{heat}:
\begin{equation}\label{SigmaA}
\Sigma_{\alpha\beta,\gamma\delta}^{X}\equiv
	\frac{X_{\alpha\beta}(B)X_{\gamma\delta}(-B)}{X_{\alpha\beta}^{\text{linear}}(B)X_{\gamma\delta}^{\text{linear}}(-B)},\qquad
A_{\alpha\beta,\gamma\delta}^{X}\equiv\frac{X_{\alpha\beta}(B)}{X_{\gamma\delta}(-B)},
\end{equation}
where $X_{\alpha\beta}$ indicates the differential transport coefficients ${\cal{G}}_{\alpha\beta}$, ${\cal{L}}_{\alpha\beta}$, ${\cal{R}}_{\alpha\beta}$, and ${\cal{K}}_{\alpha\beta}$ defined by
\begin{align}
&{\cal{G}}_{\alpha\beta}\equiv~\frac{\partial I_{\alpha}}{\partial V_{\beta}}\bigg|_{\{\theta\}=0}
	=G_{\alpha\beta}+2G_{\alpha\beta\beta}V_{\beta}+\sum_{\epsilon\ne\beta}(G_{\alpha\beta\epsilon}
		+G_{\alpha\epsilon\beta})V_{\epsilon},\\
&{\cal{L}}_{\alpha\beta}\equiv~\frac{\partial I_{\alpha}}{\partial\theta_{\beta}}\bigg|_{\{V\}=0}
	=L_{\alpha\beta}+2L_{\alpha\beta\beta}\theta_{\beta}+\sum_{\epsilon\ne\beta}(L_{\alpha\beta\epsilon}
		+L_{\alpha\epsilon\beta})\theta_{\epsilon},\label{L}\\
&{\cal{R}}_{\alpha\beta}\equiv\frac{\partial{\cal{J}}_{\alpha}}{\partial V_{\beta}}\bigg|_{\{\theta\}=0}
	=R_{\alpha\beta}+2R_{\alpha\beta\beta}V_{\beta}+\sum_{\epsilon\ne\beta}(R_{\alpha\beta\epsilon}
		+R_{\alpha\epsilon\beta})V_{\epsilon},\\
&{\cal{K}}_{\alpha\beta}\equiv\frac{\partial{\cal{J}}_{\alpha}}{\partial\theta_{\beta}}\bigg|_{\{V\}=0}
	=K_{\alpha\beta}+2K_{\alpha\beta\beta}\theta_{\beta}+\sum_{\epsilon\ne\beta}(K_{\alpha\beta\epsilon}
		+K_{\alpha\epsilon\beta})\theta_{\epsilon},
\end{align}
and $X_{\alpha\beta}^{\text{linear}}$ refers to the linear terms $G_{\alpha\beta}$, $L_{\alpha\beta}$, $R_{\alpha\beta}$, and $K_{\alpha\beta}$.
	We here consider either an isothermal, i.e., $\{\theta\}=0$, or an isoelectric case, i.e., $\{V\}=0$, hence the terms $M_{\alpha\beta\gamma}$ and $H_{\alpha\beta\gamma}$ in Eqs.~\eqref{elec} and~\eqref{heat} are not considered.
	Notice that $X_{\alpha\beta}$ contains both linear and nonlinear terms and in the linear response regime it satisfies $\Sigma_{\alpha\beta,\beta\alpha}^{X}=A_{\alpha\beta,\beta\alpha}^{X}=1$, due to the microscopic reversibility $X_{\alpha\beta}^{\text{linear}}(B)=X_{\beta\alpha}^{\text{linear}}(-B)$.
	Hence, a deviation from 1 indicates the magnetic-field asymmetry in the nonlinear regime.
	Moreover, these definitions have direct relevance to experiments~\cite{mat12} and also are related to the efficiency of the thermoelectric power generation or the refrigeration~\cite{sai11,ben11,bra13}.

\section{Quantum Hall bar wih an antidot}
\begin{figure}[htbp]
\centering
\includegraphics[width=0.5\textwidth, clip]{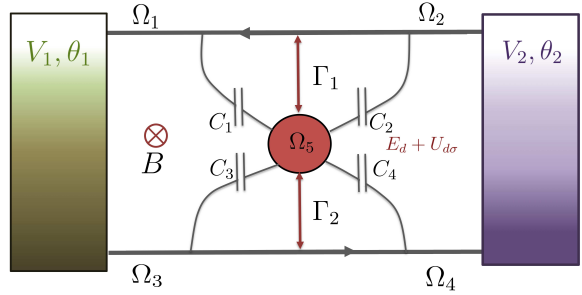}
\caption{A two-terminal quantum Hall bar-antidot system, where voltages $V_{1}$, $V_{2}$ and temperature gradients $\theta_{1}$, $\theta_{2}$ are applied.
An antidot ($\Omega_{5}$) is coupled to the edge states with the hybridization widths $\Gamma_{1}$ and $\Gamma_{2}$ and the Coulomb interactions are described by capacitances $C_{1}$, $C_{2}$, $C_{3}$, $C_{4}$.}\label{fig:QHbar}
\end{figure}

	As a model system for the general formalism described above, we consider a conductor in the quantum Hall regime in a two-terminal setup, see Fig.~\ref{fig:QHbar}.
	We suppose that $B$ is strong enough so that only the lowest Landau level is occupied and consider the reversal of its direction $B\to-B$.
	A gate-controllable antidot~\cite{for94,kir94} can connect two counter-propagating edge states, which we regard as a quantum impurity with a Breit-Wigner resonance at $E_{d}+eU_{d}(B)$ where $U_{d}(B)$ is the potential shift at the antidot in the presence of magnetic field $B$.
	The upper and lower edge states are tunnel-coupled to the antidot level via coupling strengths $\Gamma_{1}$ and $\Gamma_{2}$, respectively.
	If the direction of the magnetic field is reversed, the resonant level at the antidot is located at $E_{d}+eU_{d}(-B)$.
	It should be noted that in general $U_{d}(B)\ne U_{d}(-B)$ beyond the linear response regime~\cite{san04}.
	The $B$-asymmetry can appear either via scattering asymmetry, $\Gamma_{1}\ne\Gamma_{2}$, or electrical asymmetry if the Coulomb interaction between the upper edge and the antidot charges is stronger (or weaker) than that between the lower edge and antidot.

	In Fig.~\ref{fig:QHbar}, we divide the conductor potential into five regions $\Omega_{i}$ with $i=1,\dots,5$, with $\Omega_{5}\equiv\Omega_{d}$ indicating the antidot region.
	We assume that the potential $U_{i}$ in each region is constant and the Coulomb interactions between different regions are described by a capacitance matrix $C_{ij}$~\cite{but93}, capturing the essential physics of our interest~\cite{san04,but93,chr96}.
	For definiteness, we also assume the equal density of states for all regions, i.e., $D_{i}=D$ and the symmetric injectivities between two terminals, i.e., $D_{i\alpha}^{p,e}=D^{p,e}$ and $\Pi_{i}=\Pi$.
	Finally, we self-consistently determine the internal potential and therefore the CPs by solving the Poisson's equation explained previously.
	
	We here consider the two cases:
(i) the scattering asymmetry, i.e., $C_{i}=C$, $\Gamma_{1}=(1+\eta)\Gamma/2$, $\Gamma_{2}=(1-\eta)\Gamma/2$, 
(ii) the electrical asymmetry, i.e., $\Gamma_{1}=\Gamma_{2}$, $C_{1}=C_{2}=(1+\xi)C$, $C_{3}=C_{4}=(1-\xi)C$.
	In the above two cases, the inherent asymmetry is described with a parameter $\eta$ or $\xi$, respectively.
	We evaluate the CPs as
\begin{align}
&u_{1}(B)=u_{2}(-B)=\left\{
	\begin{array}{ll}
		\frac{1}{2}+\eta c_{\text{sc}}\\
		\frac{1}{2}+\xi c_{\text{el}}
	\end{array}\right.,\quad\quad
u_{1}(-B)=u_{2}(B)=\left\{
	\begin{array}{ll}
		\frac{1}{2}-\eta c_{\text{sc}}\\
		\frac{1}{2}-\xi c_{\text{el}}
	\end{array}\right.,\label{CPu1}\\
&z_{1}(B)=z_{2}(-B)=\frac{D^{e}}{eD^{p}}u_{1}(B),\quad\qquad
z_{1}(-B)=z_{2}(B)=\frac{D^{e}}{eD^{p}}u_{1}(-B),\label{CPz1}
\end{align}
with
\begin{equation}
c_{\text{sc}}=\bigg(2+\frac{4\pi CD^{p}\Gamma}{r(C-e^{2}\Pi)}\bigg)^{-1},\quad
c_{\text{el}}=\frac{-\pi e^{2}\Pi D^{p}C\Gamma t}{(C-e^{2}\Pi)[2\pi CD^{p}\Gamma+r(C-e^{2}\Pi)]},
\end{equation}
where $r=1-t=\Gamma_{1}\Gamma_{2}/|\Lambda|^{2}$ is the Breit-Wigner reflection ($t$: transmission) through the antidot at equilibrium, and $\Lambda=E_{F}-E_{d}-eU_{d}^{\text{eq}}+i\Gamma/2$.

	In Eq.~\eqref{CPu1}, one can easily prove the fundamental sum rule $u_{1}(B)+u_{2}(B)=u_{1}(-B)+u_{2}(-B)=1$ due to the gauge invariance.
	Importantly, the CPs are generally magnetic-field asymmetric, i.e., $u_{\alpha}(B)\ne u_{\alpha}(-B)$ and $z_{\alpha}(B)\ne z_{\alpha}(-B)$.
	We also point out the properties $u_{1}(\pm B)=u_{2}(\mp B)$ and $z_{1}(\pm B)=z_{2}(\mp B)$ in Eqs.~\eqref{CPu1} and \eqref{CPz1}, due to the chiral nature of the quantum Hall system.

	The symmetry($\Sigma$) and the asymmetry($A$) parameters defined in Eq.~\eqref{SigmaA} can be evaluated straightforwardly with the CPs in Eqs.~\eqref{CPu1} and \eqref{CPz1}.
	The general expressions of these parameters for a two-terminal quantum conductor can be found in Ref.~\cite{hwa13}, to which we also refer the readers for the detailed analysis of our quantum Hall system.
	We here briefly summarize some of the interesting results.  

	Firstly, we find all the off-diagonal asymmetry parameters $A_{\alpha\beta,\beta\alpha}^{X}$ as well as the symmetry parameter $\Sigma_{11,11}^{\cal{G}}$ for electric conductance always exhibit $B$-symmetry even in the nonlinear regime:
\begin{equation}\label{const}
\Sigma_{11,11}^{\cal{G}}=A_{12,21}^{\cal{G}}=A_{12,21}^{\cal{L}}=A_{12,21}^{\cal{R}}=A_{12,21}^{\cal{K}}=1.
\end{equation}
	This originates from the chiral property $u_{1}(\pm B)=u_{2}(\mp B)$ and $z_{1}(\pm B)=z_{2}(\mp B)$.
	In addition to this, the gauge invariance condition $\sum_{\alpha}u_{\alpha}(\pm B)=1$ gives a contribution for the derivation of $\Sigma_{11,11}^{\cal{G}}=1$ because $u_{1}(B)+u_{1}(-B)=u_{1}(B)+u_{2}(B)=1$ where the latter equality comes from the gauge invariance (see Ref.~\cite{hwa13} for the details).
	

	Secondly, we find that the symmetry parameters for $\cal{L}$(thermoelectric) and $\cal{K}$(thermal) coefficients depend on the lead indices:
\begin{subequations}
\begin{align}
&\Sigma_{11,11}^{\cal{L}}=1+c_{1}^{\cal{L}}(2\theta/T),\qquad
\Sigma_{12,21}^{\cal{L}}=1+c_{2}^{\cal{L}}(2\theta/T),\label{Sigma}\\
&\Sigma_{11,11}^{\cal{K}}=1+c_{1}^{\cal{K}}(2\theta/T),\qquad
\Sigma_{12,21}^{\cal{K}}=1+c_{2}^{\cal{K}}(2\theta/T).
\end{align}
\end{subequations}
	This different tendency between diagonal and off-diagonal parameters have a relevance to recent experiments~\cite{mat12}.
	Intriguingly, we note that this is a high-temperature effect since in the limit $k_{B}T\to0$ we find $\Sigma_{11,11}^{\cal{L}}=\Sigma_{12,21}^{\cal{L}}=\Sigma_{11,11}^{\cal{K}}=\Sigma_{12,21}^{\cal{K}}=1+2\theta/T$, irrespective of the system parameters (see Ref.~\cite{hwa13} for the details).

	In this quantum Hall system, eight parameters $\Sigma_{11,11}^{X}$ and $A_{12,21}^{X}$ for all $X={\cal{G}}$, ${\cal{L}}$, ${\cal{R}}$, ${\cal{K}}$, are independent of the scattering asymmetry($\eta$) and the electrical asymmetry($\xi$) factors due to the chiral nature, including five $B$-symmetric parameters in Eq.~\eqref{const}.
	However, the off-diagonal elements $\Sigma_{12,21}^{\cal{L}}$ and $\Sigma_{12,21}^{\cal{K}}$ are dependent on $\eta$ or $\xi$.
	Consequently, the distinction between the diagonal and the off-diagonal elements disappears for $\eta=\xi=0$.
	Thus, an asymmetry inherent in the system is crucial to observe this distinction, which is consistent with an asymmetric scattering in a recent experiment~\cite{mat12}.
	
	Even with a nonzero $\eta$ or $\xi$, our results show that we can tune the antidot level to make $c_{1}^{\cal{L}}=c_{2}^{\cal{L}}=1$ in Eq.~\eqref{Sigma} and recover the universality of the thermoelectric coefficients, i.e., $\Sigma_{11,11}^{\cal{L}}=\Sigma_{12,21}^{\cal{L}}=1+2\theta/T$.
	But, this is not the case for the heat current counterparts $\Sigma_{11,11}^{\cal{K}}$ and $\Sigma_{12,21}^{\cal{K}}$.

	In addition to the above parameters, one can also write $\Sigma_{12,21}^{\cal{G}}=1-c^{\cal{G}}(eV/\Gamma)$, $A_{11,11}^{\cal{G}}=1+c^{\cal{G}}(eV/\Gamma)$, $A_{11,11}^{\cal{L}}=1+c_{A}^{\cal{L}}(2\theta/T)$, $A_{11,11}^{\cal{R}}=1+c_{A}^{\cal{R}}(eV/\Gamma)$, and $A_{11,11}^{\cal{K}}=1+c_{A}^{\cal{K}}(2\theta/T)$, where $c^{\cal{G}}=c_{A}^{\cal{L}}=c_{A}^{\cal{R}}=c_{A}^{\cal{K}}=0$ if $\eta=\xi=0$.
	Hence $B$-asymmetry of these parameters is due only to the underlying asymmetry in the system.
	We note here that $\Sigma_{12,21}^{\cal{G}}$ and $A_{11,11}^{\cal{G}}$ are described by a single constant $c^{\cal{G}}$.
	
	Finally, we write the electrothermal symmetry parameters $\Sigma_{11,11}^{\cal{R}}=1+c_{1}^{\cal{R}}(eV/\Gamma)$ and $\Sigma_{12,21}^{\cal{R}}=1-c_{2}^{\cal{R}}(eV/\Gamma)$ in heat current measurements.
	Here, $\Sigma_{12,21}^{\cal{R}}$ is in general a function of $\eta$ or $\xi$, whereas $\Sigma_{11,11}^{\cal{R}}$ is not.

\subsection{Nonlinear Peltier effect}
When an electrical current ($I$) flows through a conductor in a isothermal configuration a heat current ($J$) is generated.
The linear Peltier coefficient is defined as the ratio between the heat and electrical currents,
\begin{equation}
\Pi_0=\frac{R_{11}}{G_{11}}\,,
\end{equation}
In Ref. \cite{lop13b}  the $\Pi$ coefficient  was generalized to the nonlinear regime.
By keeping only the leading order corrections in powers of the electrical current 
one has 
\begin{equation}\label{peltiernon}
\delta \Pi=\Pi-\Pi_0=\frac{I}{G_{11}}\left[\frac{R_{111}}{R_{11}}-\frac{G_{111}}{G_{11}}+\cdots\right]\,,
\end{equation}
The conversion factor for electric currents into heat flow in the nonlinear regime is given by the relative strength of the nonlinear conductances, $R_{111}$, and $G_{111}$ to the linear ones ($R_{11}$, and $G_{11}$). In terms of the symmetries of the nonlinear coefficients we can rewrite Eq. (\ref{peltiernon}) as

\begin{equation}
\delta \Pi=\Pi-\Pi_0=\frac{I}{2 V G_{11}}\left[\Sigma_{11,11}^{\cal R}+A_{11,11}^{\cal{R}}-\left(\Sigma_{11,11}^{\cal G}+A_{11,11}^{\cal{G}}\right)+\cdots\right]\,,
\end{equation}
\begin{figure}[htbp]
\centering
\includegraphics[angle=270,width=0.6\textwidth, clip]{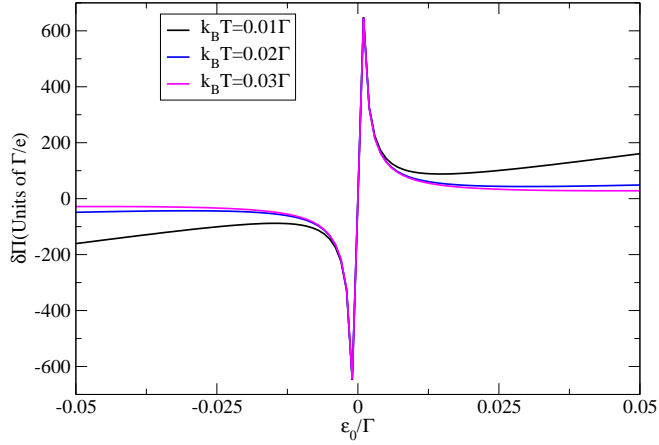}
\caption{Plots of $\delta\Pi$ versus antidot level $\varepsilon_0/\Gamma$ for several background temperatures $k_{B}T$.
We used $k_{B}\theta/\Gamma=eV/\Gamma=0.1$ and $\eta=0.1$.}\label{Fig3}
\end{figure}
In Fig. \ref{Fig3} we plot $\delta \Pi$ normalized to a given value of the electric current $I$.  We observe large deviations of the Peltier effect arising
around resonance (i.e., when $\epsilon_0\approx E_F$). Otherwise, the nonlinear Peltier coefficient attains its linear value. Besides, we find that deviations occurring for  $\epsilon_0\approx E_F$ are almost independent of the background temperature. Only when $\epsilon_0$ lies off resonance, the largest deviations take place for the highest temperature. 

\subsection{Wiedemann-Franz law}

We now explore the degree of fulfilment of the Wiedemann-Franz law beyond linear response. For a metallic system and at very low temperatures the ratio between the thermal and electric conductances normalized to the background temperatue $T$ equals a universal value,
\begin{equation}
\Lambda_0=\frac{K_{11}}{T G_{11}}\,,
\end{equation}
where $\Lambda_0= \frac{\pi^2}{3}(\frac{k_B}{e})^2$ is the Lorentz number. In the nonlinear regime, Ref. \cite{lop13b} suggests a similar quantity as the ratio between the heat flow (normalized to the temperature shift) in the isoelectrical configuration and the electrical current in the isothermal case (normalized to the applied voltage),

\begin{equation}
\Lambda=\frac{(J/\theta)|_{V=0}}{T (I/V)|_{\theta=0}}\,,
\end{equation}
The normalized deviation of the Wiedemann-Franz law from the Lorentz number can be expressed
in terms of the nonlinear transport coefficients,
\begin{equation}
\frac{\Lambda-\Lambda_0}{\Lambda_0}=\frac{K_{111}}{K_{11}} V-\frac{G_{111}}{G_{11}}\theta+\cdots
\end{equation}
which, in turn, can be recast using the symmetry and asymmetry coefficients defined above,
\begin{equation}
\frac{\Lambda-\Lambda_0}{\Lambda_0}=\left(\frac{\Sigma_{11,11}^{\cal{K}}+ A_{11,11}^{\cal{K}}}{2}-1\right) \frac{V}{2\theta}-
\left(\frac{\Sigma_{11,11}^{\cal{G}}+ A_{11,11}^{\cal{G}}}{2}-1\right) \frac{\theta}{2V}+\cdots
\end{equation}
\begin{figure}[htbp]
\centering
\includegraphics[angle=270,width=0.6\textwidth, clip]{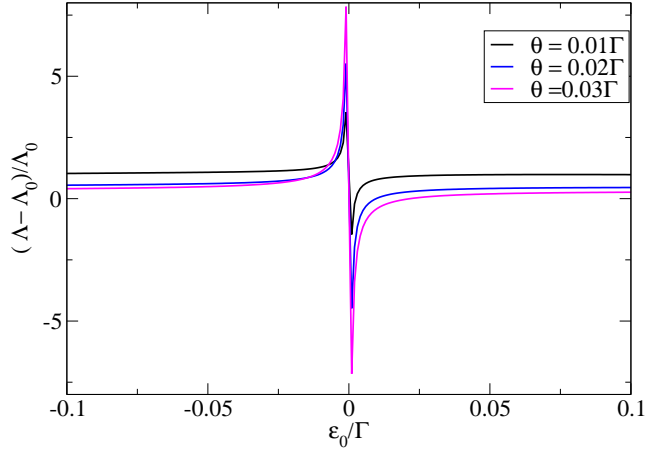}
\caption{Plots of $(\Lambda-\Lambda_0)/\Lambda_0$ versus antidot level $\varepsilon_0/\Gamma$ for several temperatures $k_{B}\theta$.
We use $k_{B}T/\Gamma=0.1$, $eV/\Gamma=0.05$ and $\eta=0.5$.}\label{Fig4}
\end{figure}

We find that departures become stronger for level resonances around the Fermi energy within a energy scale given by $\Gamma$.
This is so because the $\Gamma$ determines the energy variation of the transmission function.
In addition, the deviations increase for higher $\theta$ since the heat transport is more linear with increasing $\theta$.
		
\section{Thermoelectric transport in quantum spin Hall insulators}
\subsection{Spin-generalized screening potential}
\begin{figure}[htbp]
\centering
\includegraphics[width=0.45\textwidth, clip]{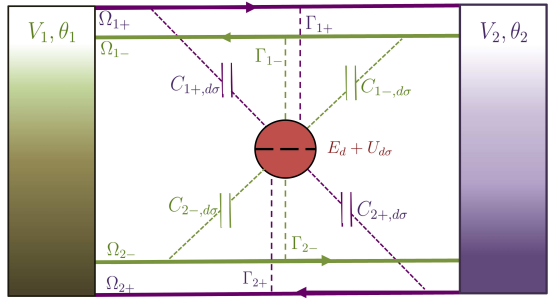}
\caption{Schematics of quantum spin Hall bar antidot setup.
Spin-generalized interactions are described by spin-dependent capacitance $C_{is,d\sigma}$,
with $i=1,2$ being the edge label, $s=\pm$ the helicity, $d$ standing for the dot, and $\sigma=\uparrow,\downarrow$ the electronic spin.
Spin-dependent couplings between the helical edges and the single level antidot are denoted with $\Gamma_{is}$. 
}\label{fig:spinhallthermal}
\end{figure}

	In the previous section, a quantum Hall antidot in a two-terminal setup has been investigated, where we determine the CPs by solving the Poisson's equation.
	Here, we consider a simple extension to the quantum spin Hall (QSH) system, see Fig.~\ref{fig:spinhallthermal}.
	In order to deal with the spintronic case with the same methodology, we generalize the scattering formalism in a spin-dependent manner.
	The potential $U_{\sigma}=U(\vec{r},\{V_{\gamma}\},\{\theta_{\gamma}\},\sigma)$ is now a function of the spin index $\sigma=\up,\down$ as well.
	This $\sigma$-dependence is crucial in our QSH system due to the underlying helicity, i.e., the spin-momentum correlation.
	Indeed, the property $U_{\up}\ne U_{\down}$ through the antidot filter is the operational principle for the spin-polarized currents even with normal metallic contacts.

	We generalize Eqs.~\eqref{I} and \eqref{J} into the spin-resolved form~\cite{hwa14}
\begin{align}	
&I_{\alpha}^{\sigma}=\frac{e}{h}\sum_{\beta}\int dE A_{\alpha\beta}^{\sigma}
(E,eU)f_{\beta}(E),\label{I_sigma}\\
&{\cal J}_{\alpha}^{\sigma}=\frac{1}{h}\sum_{\beta}\int dE(E-\mu_{\alpha}) A_{\alpha\beta}^{\sigma}(E,eU)f_{\beta}(E),\label{J_sigma}
\end{align}
for which we divide $2A_{\alpha\beta}$ in Eqs.~\eqref{I} and \eqref{J} into $A_{\alpha\beta}^{\up}=A_{\alpha\beta}(U_{\up})$ and $A_{\alpha\beta}^{\down}=A_{\alpha\beta}(U_{\down})$, explicitly incorporating the spin-dependent screening effect.
	In a two-terminal setup, we have $A_{11}^{\sigma}=A_{22}^{\sigma}=-A_{12}^{\sigma}=-A_{21}^{\sigma}=t^{\sigma}(E)$ where $t^{\sigma}(E)$ is the spin-resolved transmission probability.
	We disregard the spin-flip scattering, hence the current conservation condition is satisfied for each spin component as $\sum_{\alpha}I_{\alpha}^{\sigma}=0$ and $\sum_{\alpha}({\cal J}_{\alpha}^{\sigma}+I_{\alpha}^{\sigma}V_{\alpha})=0$.
	One can define the direction of spin-resolved currents $I_{\sigma}\equiv I_{1}^{\sigma}=-I_{2}^{\sigma}$ and ${\cal J}_{\sigma}\equiv {\cal J}_{1}^{\sigma}=-{\cal J}_{2}^{\sigma}-I_{\sigma}(V_{1}-V_{2})$, and hence the spin-polarized currents as well:
\begin{align}
I_{s}&= I_{\up}-I_{\down} \\
{\cal J}_{s}&= {\cal J}_{\up}-{\cal J}_{\down}
\end{align}
Total charge and heat fluxes are given by $I_{c}\equiv I_{\up}+I_{\down}$ and ${\cal J}_{c}\equiv {\cal J}_{\up}+{\cal J}_{\down}$, respectively.
	
	The screening potential $U=\sum_{\sigma}U_\sigma$ in Eq.~\eqref{eq:U1} can now be generalized as
\begin{equation}\label{eq:U}
U=U_{\text{eq}}+\sum_{\alpha,\sigma}u_{\alpha\sigma}V_{\alpha}+\sum_{\alpha,\sigma}z_{\alpha\sigma}\theta_{\alpha},
\end{equation}
where $u_{\alpha\sigma}=(\partial U_{\sigma}/\partial V_{\alpha})_{\text{eq}}$ and $z_{\alpha\sigma}=(\partial U_{\sigma}/\partial\theta_{\alpha})_{\text{eq}}$ are spin-dependent CPs which relate the variation of spin-resolved potential $U_{\sigma}$ to voltage and temperature shifts at each terminal.

As described earlier, the self-consistent determination of $U$ can be accomplished via Poisson's equation $\nabla^{2}\Delta U=-4\pi q$, with $\Delta U=U-U_{\text{eq}}=\sum_{\sigma}\Delta U_{\sigma}$ and
\begin{equation}\label{q}
q=\sum_{\sigma}q_{\sigma}=e\sum_{\alpha,\sigma}\Big[D_{\alpha}^{p}(\sigma)eV_{\alpha}+D_{\alpha}^{e}(\sigma)\theta_{\alpha}\Big]
	+e^{2}\sum_{\sigma}\Pi_{\sigma}\Delta U_{\sigma}\,.
\end{equation}
	In our model, $\sigma$-dependences of $D_{\alpha}^{p,e}(\sigma)$ and $\Pi_{\sigma}$ appear for unequal spin populations arising, e.g., from ferromagnetic leads.
	In order to understand this, note that the first two terms in a square bracket in Eq.~\eqref{q} are the contributions from the lead injection of charges by means of voltage and thermal driving, respectively.
	Only when the spin population is unequal ($p\ne0$, $p$: polarization) via using ferromagnetic contacts, one expects $D_{\alpha}^{p,e}(\up)\ne D_{\alpha}^{p,e}(\down)$ and $\Pi_{\up}\ne\Pi_{\down}$, where the latter comes directly from $D_{\up}\ne D_{\down}$.
	For normal metallic contacts, the only term in Eq.~\eqref{q} which can give rise to a spin imbalance inside the system is the screening potential $\Delta U_{\sigma}$.

	The solution procedure is quite analogous to the quantum Hall case (see Ref.~\cite{hwa14} for details).
	It should be noted that for QSH case the edge-antidot couplings $\Gamma_{1s}$ and $\Gamma_{2s}$ in general depend on the helicity $s=\pm$ corresponding to spin channels $\up$($+$) and $\down$($-$), when coupled to spin-polarized ferromagnetic contacts.
	This is also the case for the reflection and transmission probabilities since $r_{\sigma}=1-t_{\sigma}=\Gamma_{1s}\Gamma_{2s}/|\Lambda_{s}|^{2}$, where $\Lambda_{s}=E_{F}-E_{d}+i\Gamma_{s}/2$ with $\Gamma_{s}=\Gamma_{1s}+\Gamma_{2s}$.
	But, for normal contacts with $p=0$, there is no spin imbalance at the edge states, which leads to $t_{\up}=t_{\down}$ via antidot scattering.
	In this case, the linear conductances are spin-independent and the spin-polarization appears only in the nonlinear transport regime.
	In contrast, for general case with $p\ne0$, unequal spin density leads to $t_{\up}(p)\ne t_{\down}(p)$ giving rise to spin-polarized electric and heat currents already in the linear regime (see Ref.~\cite{hwa14} for details).
	However, even for $p\ne0$ case, if the two contacts are magnetized in an antiparallel configuration, one finds $t_{\up}(p)=t_{\down}(p)$ and hence the linear spin-polarization vanishes as in normal contacts, due to the helical nature of the QSH system.
		
	The effective Poisson's equation reads
\begin{equation}\label{Poisson}
q_{is}=e\sum_{\alpha}(D_{is,\alpha}^{p}eV_{\alpha}+D_{is,\alpha}^{e}\theta_{\alpha})+e^{2}\Pi_{is}\Delta U_{is}
		=\sum_{\sigma}C_{is,d\sigma}(\Delta U_{is}-\Delta U_{d\sigma}).
\end{equation}
	Note here that the charge with spin $\sigma=\up$($\down$) at the antidot is supplied from the edge with helicity $s=+$($-$) since in our model we neglect the spin-flip scattering, by which one can maximize spin-polarization effects.
	The density of states for all regions are given by $D_{is}=D_{s}=(1+sp)D/2$, and the symmetric injectivities from the two terminals give $D_{is,\alpha}^{p,e}=D_{s}^{p,e}=(1+sp)D^{p,e}/2$ and $\Pi_{is}=\Pi_{s}=(1+sp)\Pi/2$.

\subsection{Normal contacts}
	Here, we focus on the result for normal metallic contacts ($p=0$) since this is the most attractive case of an all-electrical setup, in which our main physics of interest resides.
	For $p=0$, it should be mentioned that the spin-channel density of states does not depend on the helicity, i.e., $D_{+}=D_{-}=D/2$ with $D=D_{+}+D_{-}$.	

	We consider the scattering asymmetric case, i.e., equal interaction strength $C_{is,d\sigma}=C_{is}=C_{s}=C/2$ with $C=C_{+}+C_{-}$, but asymmetric hybridizations $\Gamma_{1s}=(1+\eta)\Gamma/4$, $\Gamma_{2s}=(1-\eta)\Gamma/4$ with $\Gamma=\Gamma_{+}+\Gamma_{-}$ ($\Gamma_{s}=\Gamma_{1s}+\Gamma_{2s}=\Gamma/2$).
	The coupling asymmetry is thus quantified with a nonzero $\eta=(\Gamma_{1}-\Gamma_{2})/\Gamma$ where $\Gamma_{i}=\sum_{s}\Gamma_{is}$.
	In an experiment, this can be the general case where the antidot is closer to one of the edge states.
	Also, this asymmetry can be generated by tuning the width and the height of the tunnel barriers between the antidot level and the edge channels.
	One can find $\Delta U_{d\sigma}=u_{1\sigma}V_{1}+u_{2\sigma}V_{2}+z_{1\sigma}\theta_{1}+z_{2\sigma}\theta_{2}$, where the corresponding CPs are given by
\begin{align}
&u_{1\up}=u_{2\down}=\frac{1}{2}+\eta c_{\text{sc}},\quad
u_{1\down}=u_{2\up}=\frac{1}{2}-\eta c_{\text{sc}},\label{CPu}\\
&z_{1\up}=z_{2\down}=\frac{D^{e}}{eD^{p}}u_{1\up},\quad
z_{1\down}=z_{2\up}=\frac{D^{e}}{eD^{p}}u_{1\down}.\label{CPz}
\end{align}
Here, $c_{\text{sc}}=[2-2C/e^{2}\Pi]^{-1}=C_\mu/2C$ with $1/C_\mu=1/C+1/e^2D$.
Importantly, we can expect electronic transport to be spin polarized for asymmetric couplings due to the spin-dependent screening.
Intriguingly, our result is a purely interaction-driven effect and this effect disappears in the noninteracting limit $C\to\infty$.

	Up to the second order expansion of $V_{1}=V$ and $\theta_1=\theta$ ($V_2=\theta_2=0$), we have the current expressions
\begin{align}
I_{s}&=-\eta c_{\text{sc}}\bigg(\frac{2e^{3}}{h}t'V^{2}
	+\frac{2e\pi^{2}k_{B}^{2}T}{3h}\frac{D^{e}}{D^{p}}t''\theta^{2}
+\frac{2e^{2}}{h}\bigg[\frac{\pi^{2}k_{B}^{2}T}{3}t''
+\frac{D^{e}}{D^{p}}t'\bigg]V\theta\bigg),\label{eq:Is_eta}\\
\calj_{s}&=-\eta c_{\text{sc}}\bigg(\frac{2e^{2}\pi^{2}(k_{B}T)^{2}}{3h}t''V^{2}
+\frac{2\pi^{2}k_{B}^{2}T}{3h}\frac{D^{e}}{D^{p}}t'\theta^{2}
+\frac{2e\pi^{2}(k_{B}T)^{2}}{3h}\bigg[\frac{1}{T}t'+\frac{D^{e}}{D^{p}}t''\bigg]V\theta\bigg),\label{eq:Js_eta}\\
I_{c}&=\frac{2e^{2}}{h}tV
+\frac{2e\pi^{2}k_{B}^{2}T}{3h}t'\theta
		+\frac{e\pi^{2}k_{B}^{2}}{3h}\bigg(t'-T\frac{D^{e}}{D^{p}}t''\bigg)\theta^{2}
+\frac{e^{2}}{h}\bigg(\frac{\pi^{2}k_{B}^{2}T}{3}t''-\frac{D^{e}}{D^{p}}t'\bigg)V\theta,\label{eq:Ic_eta}
\end{align}
\begin{equation}
\begin{split}
\calj_{c}&=\frac{2e\pi^{2}(k_{B}T)^{2}}{3h}t'V
+\frac{2\pi^{2}k_{B}^{2}T}{3h}t\theta\\
&\quad
-\frac{e^{2}}{h}\bigg(t+\frac{\pi^{2}(k_{B}T)^{2}}{6}t''\bigg)V^{2}
+\frac{\pi^{2}k_{B}^{2}}{3h}\bigg(t-T\frac{D^{e}}{D^{p}}t'\bigg)\theta^{2}
+\frac{e\pi^{2}k_{B}^{2}T}{3h}\bigg(t'-T\frac{D^{e}}{D^{p}}t''\bigg)V\theta,\label{eq:Jc_eta}
\end{split}
\end{equation}
where $t\equiv t(E_{F})$, $t'\equiv\partial_{E}t(E)|_{E=E_{F}}$, and $t''\equiv\partial^{2}_{E}t(E)|_{E=E_{F}}$.
	These are central to our analytic results.
	The spin-polarized electronic and heat currents can be generated if $\eta c_{\text{sc}}\ne0$.
	Note also that spin-polarized currents appear in the nonlinear regime only.
\begin{figure}[htbp]
  \centering
\includegraphics[angle=270,width=0.6\textwidth, clip]{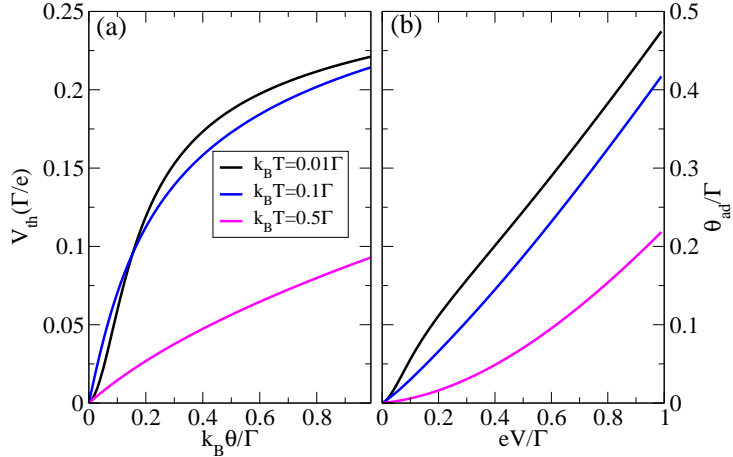}
\caption{(a) $V_{\text{th}}$ versus $k_{B}\theta/\Gamma$ and (b) $\theta_{\text{ad}}$ versus $eV/\Gamma$, at $E_{d}=0.25\Gamma$ for several $k_{B}T$. Used parameters are $\eta=c_{\text{sc}}=0.5$ and $E_{F}=0$.
}\label{Fig5}
\end{figure}
	Isothermal ($\theta=0$) and isoelectric ($V=0$) cases are treated in Ref.~\cite{hwa14}.
	Here, we focus on the pure spin currents, i.e., $I_{c}=0$, $\calj_c=0$, by means of thermoelectric Seebeck and Peltier effects.

\begin{figure}[htbp]
  \centering
\begin{tabular}{cc}
\includegraphics[angle=270,width=0.6\textwidth, clip]{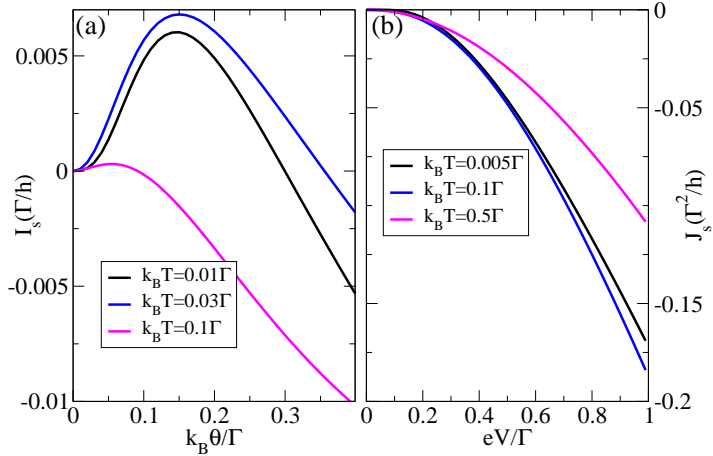}&
\end{tabular}
\caption{(a) Pure $I_s$ versus $k_{B}\theta/\Gamma$ and (b) pure $\calj_s$ versus $eV/\Gamma$, at $E_{d}/\Gamma=0.2$ for several $k_{B}T$. Here, $\eta=c_{\text{sc}}=0.5$ and $E_{F}=0$.
}\label{Fig6}
\end{figure}

	In open-circuit conditions, a thermovoltage $V_\text{th}$ can be generated in response to a temperature bias $\theta$, making $I_c=0$.
	Figure~\ref{Fig5}(a) shows the numerically evaluated set of biases $\{\theta,V\}$ which satisfies $I_{c}(V_\text{th},\theta)= 0$, where the slope can be identified as the Seebeck coefficient.
	Note that the thermovoltage acquires a nonlinear component with increasing $\theta$~\cite{lop13a,fah13}.
	One finds the pure spin current expression as a function of thermal gradient only
\begin{equation}\label{IsPure}
I_{s}=\eta c_{\text{sc}}\frac{2e\pi^{2}k_{B}^{2}T}{3h}\Bigg(\frac{\pi^{2}k_{B}^{2}T}{3}\bigg[\frac{t't''}{t}-\frac{(t')^{3}}{t^{2}}\bigg]
	+\frac{D^{e}}{D^{p}}\bigg[\frac{(t')^{2}}{t}-t''\bigg]\Bigg)\theta^{2},
\end{equation}
up to leading order in $\theta$. Figure~\ref{Fig6}(a) shows the numerical plots of pure $I_s$ versus $\theta$ beyond the quadratic
regime (a comparison with the analytical results can be found in Ref.~\cite{hwa14}). We observe that the amplitude of $I_s$ shows a nonmonotonic behavior with $T$, providing another way to maximize the effect.

In our setup, also the \textit{pure spin heat flows} can be generated using electrical means only.
By adiabatically isolating the sample, thermal bias $\theta_\text{ad}$ is generated in response to the applied voltage $V$, making $\calj_c(V,\theta_\text{ad})=0$, see Fig.~\ref{Fig5}(b).
We find the pure spin heat current expression
\begin{equation}\label{JsPure}
\calj_{s}=\eta c_{\text{sc}}\frac{2e^{2}\pi^{2}(k_{B}T)^{2}}{3h}\Bigg(\bigg[\frac{(t')^{2}}{t}-t''\bigg]\\
+T\frac{D^{e}}{D^{p}}\bigg[\frac{t't''}{t}-\frac{(t')^{3}}{t^{2}}\bigg]\Bigg)V^{2}.
\end{equation}
up to leading order in $V$. Corresponding plots are shown in Fig.~\ref{Fig6}(b) as a function of $V$.

\subsection{Polarized contacts}
	For polarized contacts with $p\ne0$, an unequal spin injection from the reservoirs with a majority spin density $(1+p)D/2$ and a minority one $(1-p)D/2$ would generate spin-polarized currents.
	In this sense, the results are to some extent anticipated and hence less interesting than those of the unpolarized case. 
	For an equal and antiparallel magnetization, however, it is interesting to note that linear conductances are spin-independent giving rise to vanishing linear spin-polarization. This will be explained below.
	
	Although the QSH bar itself preserves the time-reversal invariance, we can also investigate the reciprocity relation by considering ferromagnetic contacts with a polarization $p$, where the role of a magnetic-field reversal ($B\to-B$) can be achieved by simultaneous reversal of polarization $p\to-p$ and spin index $\sigma\to\bar\sigma$ ($\sigma=\up,\down$)~\cite{lop12}.
	As we will explain below, in any case the Onsager-Casimir relation is always satisfied for linear coefficients, i.e., $X_{\alpha\beta}^{\up}(p)=X_{\beta\alpha}^{\down}(-p)$ for $X=G,L,R,K$, which holds even for $p=0$.
	Beyond the above argument, we here discuss about the breakdown of this fundamental symmetry relation in nonlinear regime of transport for parallel and antiparallel magnetization respectively, and view the normal contacts as a limiting case of these two.

\subsubsection{Parallel magnetization}
	For an equal and parallel magnetization between two reservoirs with a polarization $p$, the density of majority spin-component (say, $\up$) is given by $D_{\ell\up}=(1+p)D_{\ell}/2$ and the minority one by $D_{\ell\down}=(1-p)D_{\ell}/2$ where $\ell=L,R$ denotes the left or right leads.
	Then, including the helicity $s$-dependence, we have $\Gamma_{\ell s}=(1+sp)\Gamma_{\ell}/2$ where $\ell=1,2$ corresponds to the upper or lower counterpart for edge-antidot coupling [see Fig.~\ref{fig:spinhallthermal}].
	This gives polarization- and spin-dependent transmission as $t^{\sigma}(p,E)=16(E-E_{d})^{2}/[16(E-E_{d})^{2}+(1+sp)^{2}\Gamma^{2}]$ with $\Gamma=\Gamma_{1}+\Gamma_{2}$.
	Hence, as expected, spin-polarized currents appear already in the linear regime since $t_{\up}(p)-t_{\down}(p)\ne0$.
	Moreover, one can notice here that $t_{\up}(p)=t_{\down}(-p)$ because the relation $sp=\bar{s}(-p)$ is always satisfied in the expression of transmission, with $s=+(-)$ corresponding to $\sigma=\up(\down)$.
	Thus, we have $G_{\alpha\beta}^{\up}(p)=G_{\beta\alpha}^{\down}(-p)$, $L_{\alpha\beta}^{\up}(p)=L_{\beta\alpha}^{\down}(-p)$, $R_{\alpha\beta}^{\up}(p)=R_{\beta\alpha}^{\down}(-p)$, and $K_{\alpha\beta}^{\up}(p)=K_{\beta\alpha}^{\down}(-p)$ for $\alpha,\beta=1,2$ (we refer the reader to Ref.~\cite{hwa14} for explicit expressions of all coefficients).
	This is a manifestation of the fundamental symmetry relations near equilibrium~\cite{but90}.

	We calculate the screening potential $\Delta U^{\sigma}(p)=u_{1}^{\sigma}(p)V_{1}+u_{2}^{\sigma}(p)V_{2}+z_{1}^{\sigma}(p)\theta_{1}+z_{2}^{\sigma}(p)\theta_{2}$, where the corresponding CPs in the presence of (i) scattering and (ii) electrical asymmetries are given by
\begin{align}
&u_{1}^{\sigma}(p)=\left\{
	\begin{array}{ll}
		\frac{1}{2}+\eta c_{\text{sc}}^{\sigma}(p)\\
		\frac{1}{2}+\xi c_{\text{el}}^{\sigma}(p)~,
	\end{array}\right.
u_{2}^{\sigma}(p)=\left\{
	\begin{array}{ll}
		\frac{1}{2}-\eta c_{\text{sc}}^{\sigma}(p)\\
		\frac{1}{2}-\xi c_{\text{el}}^{\sigma}(p)~,
	\end{array}\right.\label{CPu_p}
\end{align}
and $z_{\ell=1,2}^{\sigma}(p)=(D^{e}/eD^{p})u_{\ell=1,2}^{\sigma}(p)$, with
\begin{align}
&c_{\text{sc}}^{\sigma}(p)=s\bigg[2-\frac{2C(1+sp)}{e^{2}\Pi}\bigg]^{-1},\label{c_sc_p}\\
&c_{\text{el}}^{\sigma}(p)=\frac{-(s)e^{4}\pi\Gamma C\Pi D^{p}(1+sp)\bigg(\frac{(1+sp)^{2}(1-r_{\sigma}^{2})}{[2C-e^{2}\Pi(1+sp)(1+r_{\sigma})]^{2}}-\frac{(1+\bar{s}p)^{2}(1-r_{\bar\sigma}^{2})}{[2C-e^{2}\Pi(1+\bar{s}p)(1+r_{\bar\sigma})]^{2}}\bigg)}
{4\pi\Gamma C(1+sp)+8e^{2}r_{\sigma}}.
\end{align}
Here, $t_{\sigma}(p)=1-r_{\sigma}(p)=16(E_{F}-E_{d})^{2}/[16(E_{F}-E_{d})^{2}+(1+sp)^{2}\Gamma^{2}]$ and $s=\pm$ refers to $\sigma=\up,\down$.
	
	We note that $c_{\text{sc}}^{\up}(p)=-c_{\text{sc}}^{\down}(-p)$ and $c_{\text{el}}^{\up}(p)=-c_{\text{el}}^{\down}(-p)$, and hence $u_{1}^{\up}(p)=u_{2}^{\down}(-p)$.
	Furthermore, one can show that $u_{\ell=1,2}^{\up}(p)-u_{\ell=1,2}^{\down}(-p)\propto\eta,\xi$ and $u_{\ell=1,2}^{\up}(p)-u_{\ell=1,2}^{\down}(p)=\pm\eta[c_{\text{sc}}^{\up}(p)-c_{\text{sc}}^{\down}(p)]$ or $\pm\xi[c_{\text{el}}^{\up}(p)-c_{\text{el}}^{\down}(p)]$.
	Importantly, the former characterizes the breakdown of Onsager-Casimir symmetry in nonlinear regime~\cite{hwa13,san04}, e.g., $G_{111}^{\up}(p)-G_{111}^{\down}(-p)\propto\eta,\xi$, while the latter does the nonlinear contribution of the spin-polarized currents, e.g., $G_{111}^{\up}(p)-G_{111}^{\down}(p)=g(\eta,\xi;p)$, with $g(\eta,\xi;p)$ being a function of $\eta$ or $\xi$ and $p$.
	For $p=0$, these two relations merge into a single one $u_{1}^{\up}-u_{1}^{\down}=2\eta c_{\text{sc}},2\xi c_{\text{el}}$, leading to $G_{111}^{\up}-G_{111}^{\down}\propto\eta,\xi$. Previously, this has been shown to generate spin-polarized currents in the nonlinear transport regime with normal metallic contacts.

\begin{figure}[htbp]
  \centering
\includegraphics[angle=270,width=0.6\textwidth, clip]{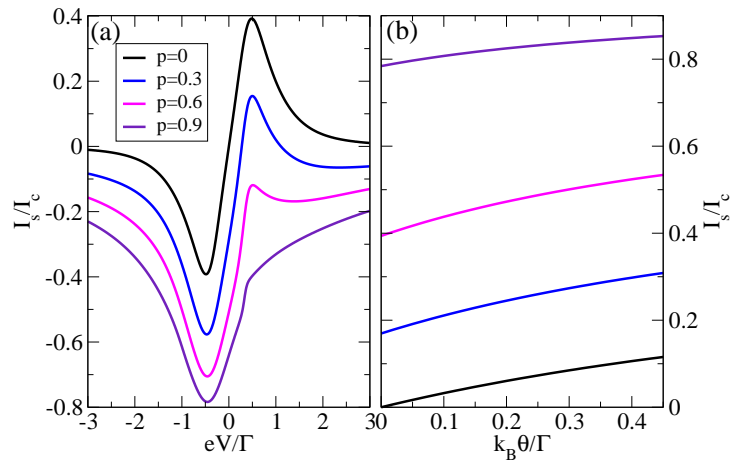}
\caption{Plots of $I_s/I_c$ versus (a) $eV/\Gamma$ at $k_BT/\Gamma=0.01$ and (b) $k_{B}\theta/\Gamma$ at $k_BT/\Gamma=0.5$, for several (parallel)  polarization $p$, with $\eta=c_{\text{sc}}=0.5$, $E_{d}/\Gamma=0.25$, and $E_{F}=0$. With $p\ne0$, spin-polarization occurs even at zero biases, i.e., $V=0$ or $\theta=0$.
}\label{Fig7}
\end{figure}

\begin{figure}[htbp]
  \centering
\includegraphics[angle=270,width=0.6\textwidth, clip]{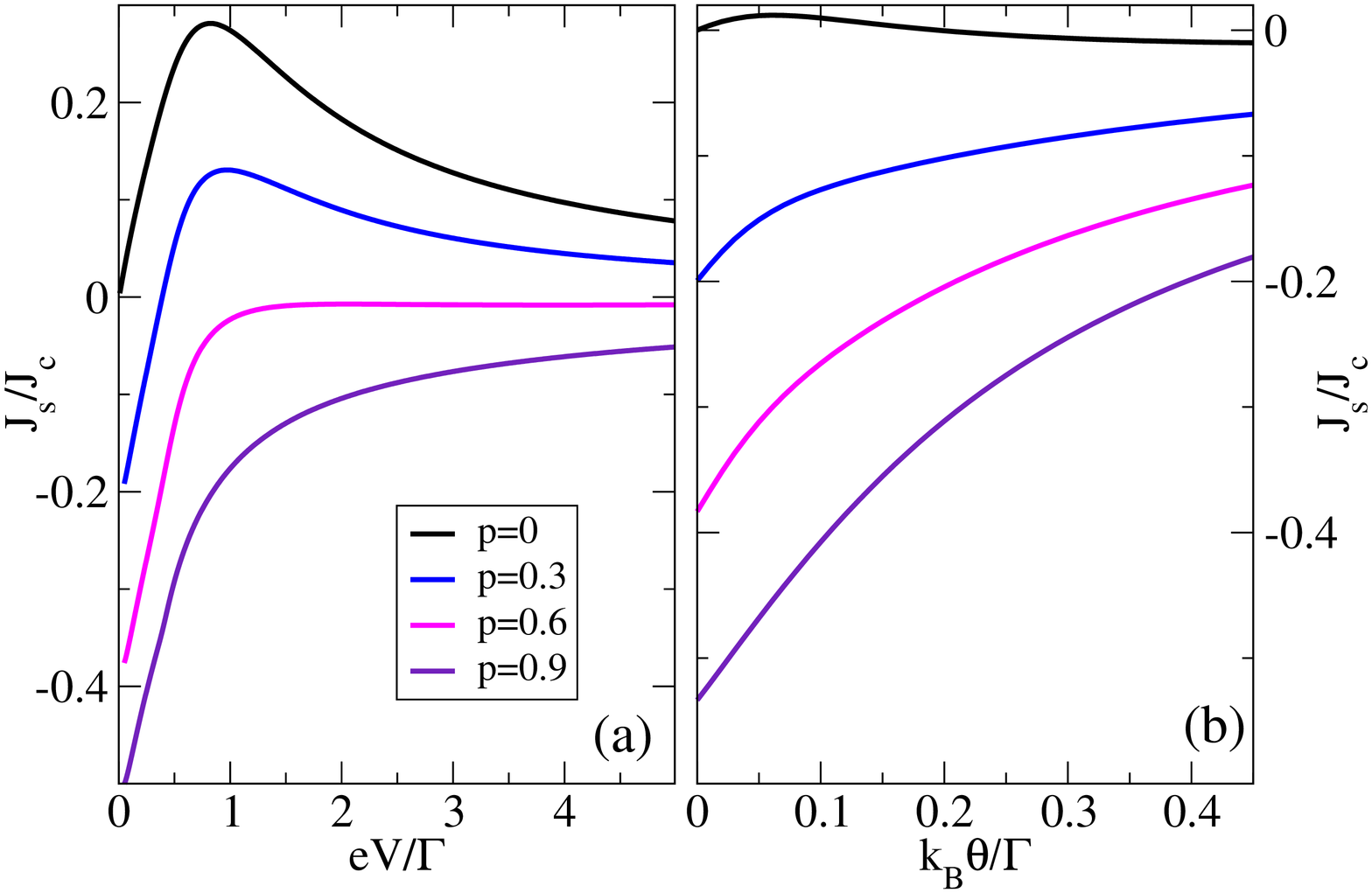}
\caption{Plots of $\calj_s/\calj_c$ versus (a) $eV/\Gamma$ at $k_BT/\Gamma=0.01$ and (b) $k_{B}\theta/\Gamma$ at $k_BT/\Gamma=0.05$, for several (parallel) polarization $p$, with $\eta=c_{\text{sc}}=0.5$, $E_{d}/\Gamma=0.3$, and $E_{F}=0$. With $p\ne0$, spin-polarization occurs even at zero biases, i.e., $V=0$ or $\theta=0$.
}\label{Fig8}
\end{figure}

\subsubsection{Antiparallel magnetization}
	For an equal but antiparallel magnetization between two reservoirs, however, the upper edge is occupied by right-moving $\up$- and left-moving $\down$-electrons both with an equal portion $(1+p)/2$ of the total density of states in each lead, because in this configuration the majority spin component is up (down) for the left (right) reservoir.
	In contrast, the lower helical edge is filled only with a minority portion $(1-p)/2$ for both left-moving [$+(\up)$] and right-moving [$-(\down)$] channels.
	The net effect of this antiparallel configuration gives spin-independent coupling and transmission, i.e., $\Gamma_{1s}=(1+p)\Gamma_{1}/2$, $\Gamma_{2s}=(1-p)\Gamma_{2}/2$, and $t^{\sigma}(p,E)=[16(E-E_{d})^{2}+p^{2}\Gamma^{2}]/[16(E-E_{d})^{2}+\Gamma^{2}]$.
	We then have $t_{\up}(p)=t_{\down}(p)$, hence linear conductance coefficients make no contribution to spin-polarized currents, i.e., $G_{\alpha\beta}^{\up}(p)=G_{\alpha\beta}^{\down}(p)$, $L_{\alpha\beta}^{\up}(p)=L_{\alpha\beta}^{\down}(p)$, $R_{\alpha\beta}^{\up}(p)=R_{\alpha\beta}^{\down}(p)$, and $K_{\alpha\beta}^{\up}(p)=K_{\alpha\beta}^{\down}(p)$.
	This vanishing linear order contribution even with a nonzero $p$ but with a specific antiparallel configuration is due to the helical nature of the QSH system.
	Nevertheless, the symmetry relation still holds, i.e., $t_{\up}(p)=t_{\down}(-p)$, hence $G_{\alpha\beta}^{\up}(p)=G_{\beta\alpha}^{\down}(-p)$, $L_{\alpha\beta}^{\up}(p)=L_{\beta\alpha}^{\down}(-p)$, $R_{\alpha\beta}^{\up}(p)=R_{\beta\alpha}^{\down}(-p)$, and $K_{\alpha\beta}^{\up}(p)=K_{\beta\alpha}^{\down}(-p)$

	We analogously evaluate $\Delta U^{\sigma}(p)=u_{1}^{\sigma}(p)V_{1}+u_{2}^{\sigma}(p)V_{2}+z_{1}^{\sigma}(p)\theta_{1}+z_{2}^{\sigma}(p)\theta_{2}$, where
\begin{align}
&u_{1}^{\sigma}(p)=\left\{
	\begin{array}{ll}
		\frac{1}{2}+c_{\text{sc}}^{\sigma}(\eta, p)\\
		\frac{1}{2}+c_{\text{el}}^{\sigma}(\xi, p)~,
	\end{array}\right.
u_{2}^{\sigma}(p)=\left\{
	\begin{array}{ll}
		\frac{1}{2}-c_{\text{sc}}^{\sigma}(\eta, p)\\
		\frac{1}{2}-c_{\text{el}}^{\sigma}(\xi, p)~,
	\end{array}\right.\label{CPu_p_anti}
\end{align}
and $z_{\ell=1,2}^{\sigma}(p)=(D^{e}/eD^{p})u_{\ell=1,2}^{\sigma}(p)$, with
\begin{align}
&c_{\text{sc}}^{\sigma}(\eta, p)=s(\eta+p)\bigg[2(1+\eta p)-\frac{2C(1-p^{2})}{e^{2}\Pi}\bigg]^{-1},\label{c_sc_anti}\\
&c_{\text{el}}^{\sigma}(\xi, p)=sp\bigg[2-\frac{2C(1-p^{2})(1+\xi p)}{e^{2}\Pi}\bigg]^{-1}.
\end{align}
Here, $t(p)=1-r(p)=[16(E_{F}-E_{d})^{2}+p^{2}\Gamma^{2}]/[16(E_{F}-E_{d})^{2}+\Gamma^{2}]$ and $s=\pm$ corresponds to $\sigma=\up,\down$.

	It is easy to show that $c_{\text{sc}}^{\up}(\eta, p)=-c_{\text{sc}}^{\down}(\eta, p)$, $c_{\text{el}}^{\up}(\xi, p)=-c_{\text{el}}^{\down}(\xi, p)$, $c_{\text{sc}}^{\up}(\eta=0, p)=c_{\text{sc}}^{\down}(\eta=0, -p)$, $c_{\text{el}}^{\up}(\xi=0, p)=c_{\text{el}}^{\down}(\xi=0, -p)$, $c_{\text{sc}}^{\sigma}(\eta=0, p)\propto sp$, $c_{\text{el}}^{\sigma}(\xi, p)\propto sp$, $c_{\text{sc}}^{\sigma}(\eta, p=0)\propto\eta s$, and $u_{1}^{\sigma}(p)=u_{2}^{\bar\sigma}(p)$.
	Thus, we have, e.g., $u_{1}^{\up}(p)-u_{1}^{\down}(-p)=c_{\text{sc}}^{\up}(\eta, p)-c_{\text{sc}}^{\down}(\eta, -p)=c_{\text{sc}}^{\up}(\eta, p)+c_{\text{sc}}^{\up}(\eta, -p)$ as well as $u_{1}^{\up}(p)-u_{1}^{\down}(p)=c_{\text{sc}}^{\up}(\eta, p)-c_{\text{sc}}^{\down}(\eta, p)=2c_{\text{sc}}^{\up}(\eta, p)$.
	
	The relation such as $u_{1}^{\up}(p)-u_{1}^{\down}(-p)=c_{\text{sc}}^{\up}(\eta, p)+c_{\text{sc}}^{\up}(\eta, -p)$ indicates the Onsager-Casimir symmetry breaking in nonlinear regime, which vanishes for $\eta=0$ but survives for $p=0$, $\eta\ne0$ with the latter corresponding to the normal metal leads.
	The nonlinear spin-polarization term, e.g., $u_{1}^{\up}(p)-u_{1}^{\down}(p)=2c_{\text{sc}}^{\up}(\eta, p)$ remains finite provided that either of $\eta$ or $p$ is nonzero.
	For $p=0$, as in the  parallel configuration, these two relations merge into a single one, e.g., $u_{1}^{\up}-u_{1}^{\down}=2\eta c_{\text{sc}},2\xi c_{\text{el}}$.
	Only when $\eta,\xi=p=0$, all spin-polarized currents vanish, which again explains the spin-filter effect in an unpolarized case [Eqs.~\eqref{eq:Is_eta} and \eqref{eq:Js_eta}], as a limiting case $p\to0$ of the antiparallel configuration.

	From the symmetry arguments in this section, one can notice that the Onsager-Casimir symmetry breakdown in nonlinear regime for $p\to0$ with a nonzero asymmetry factor $\eta$ or $\xi$ suggests the underlying principle for the observed spin-polarization for QSH antidot system.

\section{Conclusions}
We have analysed the nonlinear transport of quantum Hall setups subjected to the voltage and temperature biases. Our theory is based on the scattering transport formalism that incorporates electron-electron interaction within a mean-field description. The potential landscape of the nanostructure  depends on the injected charges due to both, the voltage and temperature gradients. Using this theory we  find the symmetry relations of the nonlinear transport coefficients for a quantum Hall bar with an inserted antidot. We find large deviations for the Wiedemann-Franz law and  Peltier coefficient from their linear values. We also have reported results for the spin Hall bar including the antidot system in a spin-dependent capacitative model. We have found pure spin current generation for which the spin energy and spin current  values depend on the magnitude of the lead polarization. 
Our results are relevant in view of recent works that emphasize nonlinear properties of thermopower in nanostructures \cite{boe01,don02,aze12,kul94,bog99,seg05,cha06,ruo11,kuo01,kra07,her13,zim14}. In general, the nonlinear regime of transport has not been explored to a large extent. Therefore, we expect the coming years to be full of exciting discoveries and  developments in this field.

\ack
Work supported by MINECO Grant No. FIS2011-23526,
the Conselleria d'Educaci\'o, Cultura i Universitats
(CAIB) and FEDER.

\end{document}